\documentstyle[aps,twocolumn,psfig]{revtex}

\begin{document}

\newcommand{\be}{\begin{equation}}
\newcommand{\ee}{\end{equation}}

\twocolumn[\hsize\textwidth\columnwidth\hsize\csname@twocolumnfalse\endcsname

\title{\bf SQUID based resistance bridge for shot noise measurement on low impedance samples}

\author{X. Jehl, P. Payet-Burin, C. Baraduc, R. Calemczuk and M. Sanquer.
\\ \small \it DSM-DRFMC-SPSMS, CEA-Grenoble,  
\\ \small \it 38054 Grenoble cedex 9, France.}

\maketitle

\begin{abstract}

We present a resistance bridge which uses a Superconducting Quantum Interference Device (SQUID) to measure the shot noise in low impedance samples ($<1\Omega$). The experimental requirements are high DC bias currents ($\lesssim 10mA$), in order to obtain sufficiently high bias voltages, together with high AC sensitivity ($\approx pA/\sqrt{Hz}$) to detect small changes of the noise away from the thermal noise ground level. These conditions are fulfilled by changing continuously the overall gain of the SQUID electronics and separating the DC and AC feedbacks. This system is used to investigate the shot noise in mesoscopic samples made with superconducting (S) and normal (N) metals ($R_{4.2K}\approx 0.25 \Omega $). This value of source resistance is out of the range of classical noise measurement schemes. Because of its low intrinsic noise ($\lesssim 5pA/\sqrt{Hz}$) our system is dominated by the thermal noise of the resistors composing the bridge ($\approx 24pA/\sqrt{Hz}$ at $4.2K$); therefore this method greatly simplifies the analysis of shot noise results. 
\\
\\
\end{abstract}
]
\section{INTRODUCTION}

\par Recent developments in mesoscopic physics revived the interest in shot noise \cite{nato,samina,martin}. Shot noise refers to the current fluctuations due to the discretness of the electron charge and was discovered by W. Schottky in 1918 in thermionic vacuum tubes. In these devices the electrons can be totally uncorrelated: in that case their emission follows a Poisson distribution and gives rise to the power spectral density:  
\be
P=2e^{*}I=P_{Poisson}  
\ee
where $e^{*}$ and $I$ are respectively the carriers charge and the net average current. The key point is that shot noise reflects the degree of correlation between electrons. In samples small enough to experience no inelastic scattering, some interesting features such as the so-called "quantum suppression of shot noise" below $P_{Poisson}$ were predicted and experimentally verified \cite{kumar}. When superconducting reservoirs are used to connect these small devices, the shot noise should increase above the prediction for normal materials because of Cooper pairs of charge $e^{*}=2e$ participating in the current transport \cite{dejong}. Experimental results regarding the interplay of shot noise suppression and Andreev reflections at NS interfaces are difficult to obtain. Most of the results were obtained with room temperature electronics with two independant FET amplifiers and cross correlation techniques \cite{glattli}. This technique is limited to source resistance of the order of $100\Omega$ at $4.2K$.
The cross correlation technique can still work with samples using a nanowire \cite{strunk} or pinholes in an insulator \cite{dielemann} as the normal metal but not with wider normal sections. We present a new experiment allowing the measurement of the shot noise in low impedance devices with high reliability since the noise of the measurement apparatus is small compared to the ground noise of interest. We will describe the low temperature resistance bridge with the SQUID first. Then we will give an overview of the two feedback loops and discuss the performances of the system.

\section{RESISTANCE BRIDGE}

\par
\begin{figure}[h]
\psfig{figure=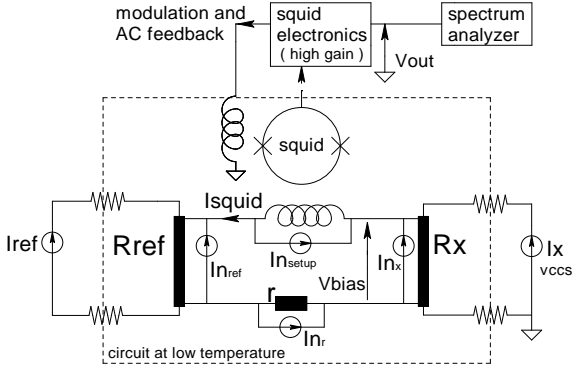,width=80mm}
\caption{\footnotesize Schematic of the circuit in the noise measurement configuration. The noise current $I_{n_{x}}$ of the sample is related to the output measured voltage noise by equation[\ref{vnout}]. At $4.2K$, $R_{ref}=0.177\Omega$, $r=4m\Omega$ and $0.25\Omega<R_{x}<0.3\Omega$. The current sources $i_{n_{ref}}$, $i_{n_{x}}$ and $i_{n_{r}}$ are the equivalent noise current generators for $R_{ref}$, ${R_{x}}$ and $r$. The noise current generator $i_{n_{setup}}$ represents the total noise of the experimental setup expressed as a current in the input coil. It is mainly due to the Voltage Controlled Current Source (VCCS) and the SQUID electronics.}
\label{fig:grandg}
\end{figure}

 We wish to perform measurements from the thermal noise regime ($V_{bias}=0$) to the full shot noise regime ($e^{*}V\gg kT$). The maximum bias current which can be applied is determined by the gap voltage of the metal used for the superconducting electrodes. For a junction made of niobium this value is about $3mV$, therefore if its resistance is $0.25\Omega$ the corresponding bias current is around $10mA$. On the other hand the ground level of the noise current is in the tens of $pA/\sqrt{Hz}$ range, so a very good resolution is required. A classical method of applying a bias voltage at low temperature consists in a reference resistance of value much smaller than $R_{x}$ placed in parallel with the latter and thus creating a low impedance voltage source \cite{devoret}. This scheme is not applicable in our case as most of the current flows through the small resistance and the $10mA$ required in the sample would impose excessively high total currents not compatible with low noise sources. We have designed a new experiment based on a resistance bridge to obtain a high sensitivity while supplying high dc bias currents. 
\\The low temperature part of the circuit consists in the resistance bridge and the SQUID in a calorimeter (Figure \ref{fig:grandg}). It is composed with a reference resistance $R_{ref}$ and the sample $R_{x}$ both connected with 4 wires. The sample is connected with short gold wires which give an additionnal resistance $r$ much smaller than $R_{ref}$ and $R_{x}$. The input coil of the SQUID is connected in series with the voltage leads by means of superconducting wires. The aim is to apply a high bias current that is not measured by the SQUID, so that the SQUID electronics can be set to a high sensitivity in order to detect small currents.
A DC current $I_{ref}$ injected through $R_{ref}$ sets the voltage $V_{bias}=R_{ref}I_{ref}$. The bridge is balanced with a DC feedback current $I_{x}$ injected in $R_{x}$ so that $i_{squid}=0$. The equation which rules the DC balance is:
\be
V_{bias}=R_{ref}I_{ref}=R_{x}I_{x}
\label{equilibre}
\ee
\\As $i_{squid}$ is then null in the DC limit, the SQUID can measure with a high gain the AC variations of $i_{squid}$ arising from  the noise of the resistances composing the bridge. The noise voltage $V_{n_{out}}$ measured by the spectrum analyzer at the output of the SQUID electronics is:
\begin{eqnarray}
V_{n_{out}}^{2}={1\over{G}^{2}}\Bigg[\left(1\over{\Sigma R}\right)^{2}\Big[{(R_{ref}i_{n_{ref}})}^{2}+{(ri_{n_{r}})}^{2}\nonumber\\
+{(R_{dyn_{x}}i_{n_{x}})}^{2}\Big]+{i_{n_{setup}}}^{2}\Bigg]
\label{vnout}
\end{eqnarray}
where $\Sigma R=r+{R_{ref}+R_{dyn_{x}}}$ and $G$ is the overall gain of the SQUID electronics $G={i_{squid}\over{V_{out}}}$. The term $R_{dyn_{x}}$ denotes the dynamic resistance of the sample which is not strictly ohmic: $R_{dyn_{x}}={\left( \partial V \over{\partial I} \right)_{I_{x}}}$. The second term in equation[\ref{vnout}] is the sum of the noise contributions of the three resistances and the non fundamental noise arising from the experimental setup expressed as a current in the input coil and noted $i_{n_{setup}}$. This equation is simply derived from the expression of $i_{squid}$ using current division of each equivalent noise current in its own source resistance and the rest of the circuit (see Figure[\ref{fig:grandg}]).

\begin{figure}[h]
\psfig{figure=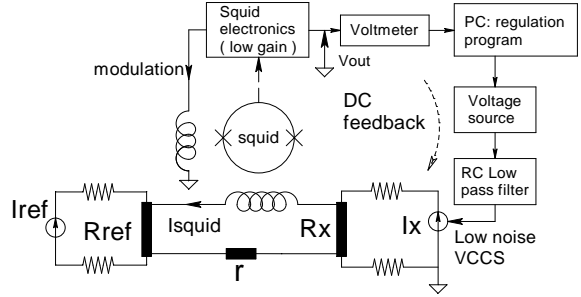,width=80mm}
\caption{\footnotesize Schematic of the circuit showing the DC feedback loop. The low noise Voltage Controlled Current Source (VCCS) is driven by the squid electronics through a regulation program. It provides the DC current $I_{x}$ which counterbalances $I_{ref}$ so that $V_{bias}$ is applied on the sample $R_{x}$. }
\label{fig:petitg}
\end{figure} 
The wiring and the temperature control are very similar to those used in reference\cite{plomb}. The SQUID and the reference resistance (made of constantan, $R_{ref}=0.177\Omega$) are kept at helium temperature by copper rods ending in the helium bath. The sample holder allows very precise temperature control. Special attention is paid to RF filtering to reduce the noise on the SQUID. We also use magnetic shielding with superconducting materials. The twisted superconducting wires connected to the SQUID are shielded with superconducting tinned tubes. Thermalization pads and reference resistance are placed in soldered lead box. Finally the whole calorimeter is surrounded by two layers of magnetic shielding tape \cite{vacuum} and a thick soldered lead foil. Flux jumps are totally suppressed and the SQUID can be operated with DC currents.

\section{FEEDBACK SCHEMES}

\par We use a Quantum Design \cite{qd} DC-SQUID and an electronics designed and built in our laboratory. It works as a Flux Locked Loop (FLL) system with modulation at $500kHz$. The FLL operation is realized on the modulation/feedback coil. 
A practical experiment includes two different stages. First the desired bias current is reached. The second stage is devoted to the noise measurement with the spectrum analyzer (SR785). 

During the first stage the current $I_{ref}$ delivered by a floating battery is increased manually and the feedback   current $I_{x}$ is fed through $R_{x}$ in order to satisfy equation \ref{equilibre}. Figure \ref{fig:petitg} shows this "DC feedback" loop. At this point the SQUID electronics has a low gain corresponding to $\approx 100{\Phi_{0}}/V$ i.e. $G\approx 20\mu A/V$. The dynamics of the system are then substantial enough to inject relatively high currents. A Labview program running on a PC regulates $V_{out}$ to keep it null while $I_{ref}$ is manually increased. First a voltmeter (Keithley2000) reads $V_{out}$; the program then drives a voltage source (HP3245A) which output is filtered by a first order $RC$ low pass filter with a long time constant. This DC feedback voltage finally drives a purposely designed low noise Voltage Controlled Current Source (VCCS).  The transconductance of this VCCS is $1mA/V$. Its output feedback current $I_{x}$ is reinjected through $R_{x}$. 
The high value of the transconductance makes it necessary to filter dramatically the noises generated by the electronics before the VCCS. Indeed input voltage noises of a few $nV/\sqrt{Hz}$ must be obtained in order to generate no more than a few $pA/\sqrt{Hz}$ at the output of the VCCS. That is why a RC low pass filter with $4.7s$ time constant is inserted before the VCCS. Therefore the only white noise generated by the DC feedback loop is the intrinsic noise of the VCCS. The VCCS is supplied with two lead batteries ($\pm 12V$) and is placed with the other lead battery providing $I_{ref}$ in a $\mu -metal$ tube closed at one end. This results in spectra showing strictly no peaks (Fig. \ref{fig:spectra}) and helps to obtain a very high stability of the SQUID electronics. 
\\When the desired bias current is reached and the bridge is balanced according to equation \ref{equilibre}, $i_{squid}$ becomes zero in the DC limit, then the gain of the SQUID electronics can be increased continuously to $1{\Phi_{0}}/V$, i.e. $G=184nA/V$. At this stage corresponding to Figure \ref{fig:grandg} the FFT analyzer can measure the noise spectrum of interest. The FFT is typically computed in the $[16Hz - 12.8kHz]$ range with 800 FFT lines. The dispersion on the spectra is $\approx 0.4pA/\sqrt{Hz}$ after 3000 averagings, the acquisition time is then less than 4 minutes. The same operations are repeated to get spectra at different bias currents.

\section{PERFORMANCE OF THE SYSTEM}

Experiments have been carried out on macroscopic metallic resistors to test the absence of shot noise due to the electronics or the reference resistor. The sample was simply replaced by a resistance made with the same constantan wire than $R_{ref}$ and of similar value. Spectra were performed from zero bias current to the mA range. They were all found superimposed within the dispersion determined by the averaging. Neither the $1/f$ noise nor the white noise were affected by the DC bias current. This indicates that no excess noise is added by the electronics or $R_{ref}$ when a bias current is supplied. Consequently any detectable current dependence of the noise observed with a real sample will undoubtedly be due to the sample itself. Also the results were constantly found to be capable of being reproduced even after warming up the experiment.

\begin{figure}
\psfig{figure=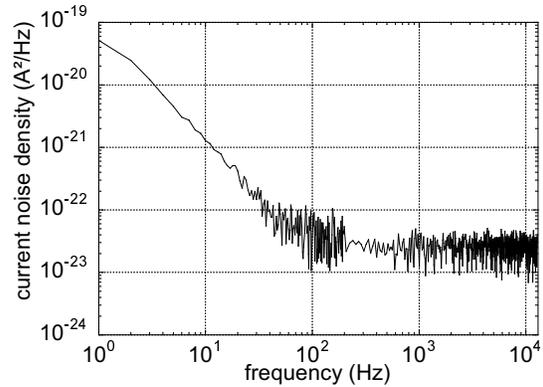,width=80mm}
\caption{\footnotesize Intrinsic total noise of the experiment. The white noise level of $2.6\times 10^{-23} A^{2}/Hz{}(5.1pA/\sqrt{Hz})$ is due to the VCCS and the SQUID system. The low frequency excess noise is due to the amplifiers composing the VCCS.} 
\label{fig:iexp}
\end{figure} 

\begin{figure}
\psfig{figure=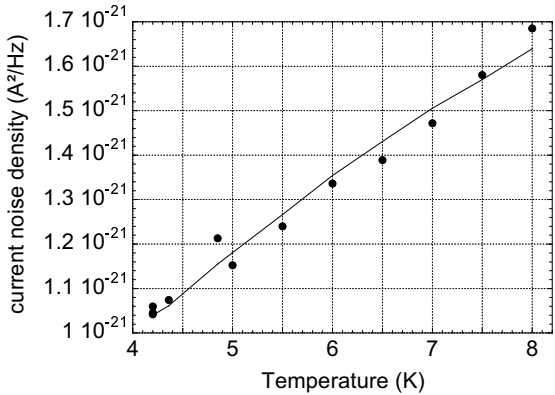,width=80mm}
\caption{\footnotesize Temperature dependance of thermal noise for a $0.22\Omega$ sample. The experimental results (dots) are very much in agreement with the Johnson-Nyquist formula (line). The latter does not yield to a straight line because the resistance is slightly temperature dependent.} 
\label{fig:johnson}
\end{figure}

Equation \ref{vnout} gives the total output noise which is measured. At zero bias current it is the sum of the thermal noise generated by the resistance bridge and the equivalent noise $i_{n_{setup}}^2$ of the whole electronics in the input coil
. As there is no shot noise at zero bias, the contribution from the three resistances is simply the thermal noise of each, given for a resistance of value R at temperature T by the Johnson-Nyquist relation $i_{n}^{2}=4k_{B}T/R$. Therefore $i_{n_{setup}}$ can be obtained from the raw datas at zero bias using equation \ref{vnout}:
\be
{i_{n_{setup}}}^{2}={(GV_{n_{out}})}^{2}-{{{4k_{B}}\over{{(\Sigma R)}^{2}}}} \left[T_{x}(R_{dyn_{x}}+r)+4.2R_{ref}\right]
\label{iexp}
\ee
The noise obtained is shown in Figure \ref{fig:iexp}. It is fitted by a constant added to a power law of the frequency to take into account the low frequency excess noise. Typically we found $i_{exp}^2=2.6\times 10^{-23}+3\times 10^{-20}/f^{1.5}$, i.e. a white noise level of $5.1pA/\sqrt{Hz}$. This noise comes from the VCCS and the SQUID setup. It is in very good agreement with noise simulations of the VCCS using SPICE modelling and with measurements of the SQUID system noise.
\\The VCCS is composed with two ultra low noise ($LT1028$ and $LT1128$) operationnal amplifiers and 4 matched resistances. A noise level of $4.5pA/\sqrt{Hz}$ is predicted by SPICE simulations using the typical noise specifications of the operationnal amplifiers and the actual values of the resistors fixing the transconductance. The second source of noise is the SQUID and its electronics which has been measured: it contributes for $\approx 2.5pA/\sqrt{Hz}$ (i.e. $13\mu\Phi_{0}/\sqrt{Hz})$. Therefore at $4.2K$ the intrinsic noise power of the experiment is only 6 \% of the thermal noise of the resistance bridge ($0.4\Omega$). 
\\To conclude let us show some results illustrating the capacity of our setup to detect small changes of noise. We can easily measure the evolution of thermal noise with temperature of a $0.22\Omega$ sample and check that it follows the Johnson Nyquist relation (Figure \ref{fig:johnson}).

\begin{figure}
\psfig{figure=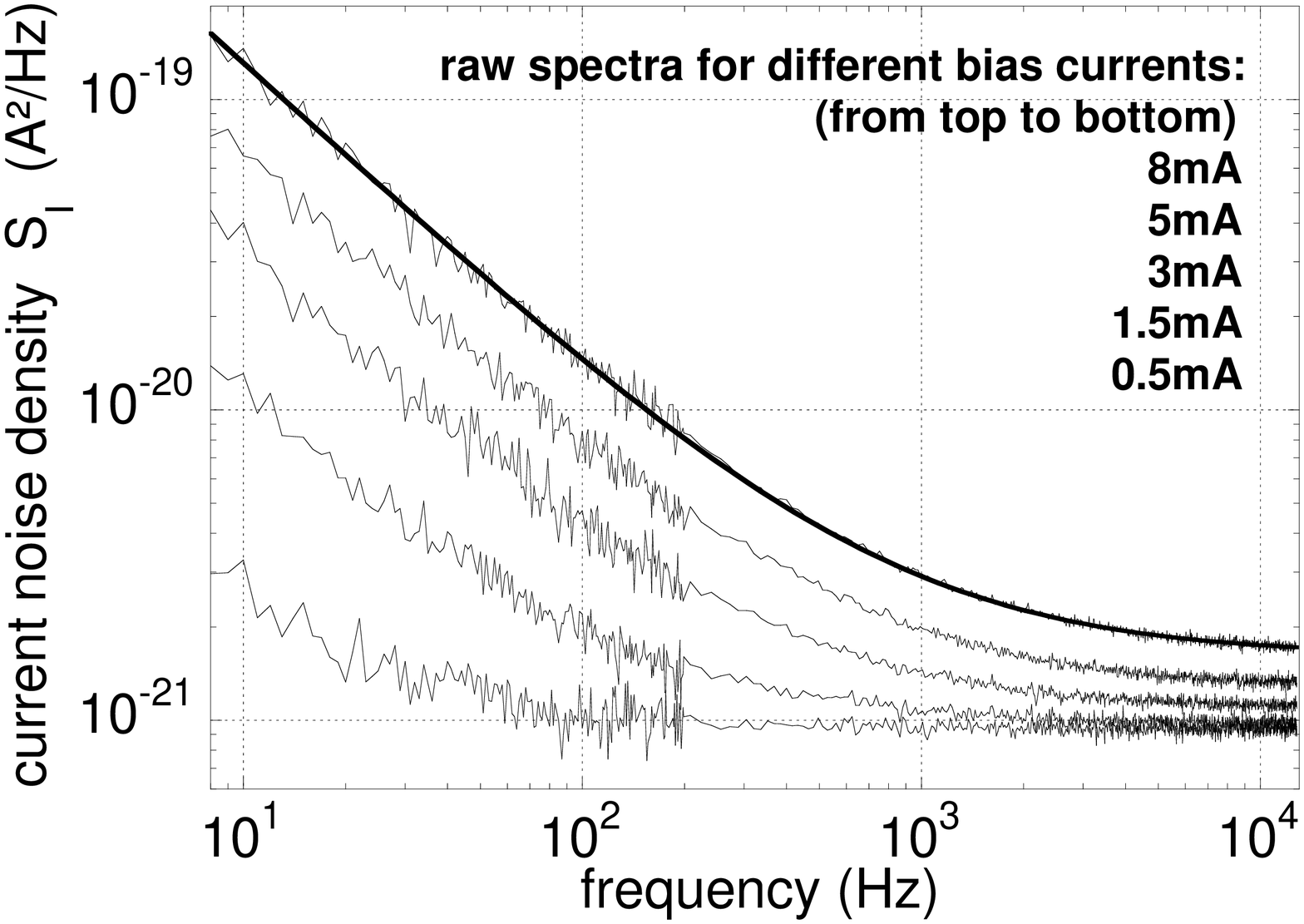,width=80mm}
\caption{\footnotesize Typical examples of $S_{I}(f)$ in a SNS junction of $\approx 0.25\Omega$ at $4.2K$. Solid curve represents fit to the $8mA$ data by the sum of a $1/f$ and frequency independant contribution. The acquisition time is about 4 minutes for the $12.8kHz$ span with 800 points, it  yields to a dispersion of $\approx 0.4pA/\sqrt{Hz}$. These data clearly show an increase of the noise level with the DC bias current.} 
\label{fig:spectra}
\end{figure} 

Because of the clear domination of the thermal noise of the bridge at zero bias, it is possible to study the crossover between the thermal and the shot noise regimes as well as the full shot noise regime . Figure \ref{fig:spectra} shows the first results obtained in an SNS (Nb-Al-Nb) junction of $\approx 0.5\mu m$ length \cite{jehl}. 
Unlike the preliminary experiment with a macroscopic metallic resistance instead of the sample, we clearly observe a modification of the noise spectra with the applied current. The low frequency noise behaves as 1/f with a coefficient roughly proportionnal to ${I_{bias}}^{2}$ indicating resistance fluctuations. The higher frequency noise which is the noise of interest clearly increases with the current. Because of its very low noise our experiment provides very reliable shot noise results with high dc bias currents. It opens new perspectives in the very rich field of shot noise measurements in mesoscopic physics.

\end{document}